\documentclass[galaxies,article,submit,pdftex,moreauthors]{Definitions/mdpi}
\firstpage{1}  
 \pubvolume{1} \issuenum{1} \articlenumber{0}
\pubyear{2025} \copyrightyear{2025}
\datereceived{ }
\daterevised{ } 
\dateaccepted{ }
\datepublished{ }
\hreflink{https://doi.org/} 

\usepackage{aas_macros}

\usepackage{amssymb}

\Title{Stellar wind parameters of  massive stars in
accretion-powered High Mass X-ray Binary Pulsars}

\TitleCitation{Stellar wind parameters of  massive stars in
accretion-powered High Mass X-ray Binary Pulsars}

\Author{Nina Beskrovnaya $^{1, 2}$*\orcidA{}, Nazar Ikhsanov $^{1,
3}$\orcidB{} and Vitaliy Kim   $^{1,4}$\orcidC{}}

\AuthorNames{Nina Beskrovnaya, Nazar Ikhsanov and Vitaliy Kim}

\isAPAStyle{%
       \AuthorCitation{Beskrovnaya, N., Ikhsanov, N., \& Kim, V.}
         }{%
        \isChicagoStyle{%
        \AuthorCitation{Beskrovnaya, Nina, Nazar Ikhsanov, and Vitaliy Kim.}
        }{
        \AuthorCitation{Beskrovnaya, N.; Ikhsanov, N,; Kim, V.}
        }
}

\address{%
$^{1}$ \quad Central Astronomical Observatory of the Russian Academy
of Sciences
at Pulkovo,  Pulkovskoe Shosse 65-1, St.\,Petersburg 196140, Russia; Beskrovnaya@yahoo.com\\
$^{2}$ \quad Special Astrophysical Observatory,  Russian Academy of
Sciences,
              Nizhnii Arkhyz, 369167 Russia;\\
$^{3}$ \quad The Institute of Applied Astronomy,  Russian Academy of
Sciences,
 Kutuzov Emb., 10, St.\,Petersburg, 191187 Russia; Nazar.Ikhsanov@gmail.com\\
$^{4}$  \quad Fesenkov Astrophysical Institute, Observatory 23,
Almaty 050020, Kazakhstan}

\corres{Correspondence: beskrovnaya@yahoo.com}

\abstract{The process of mass-exchange between the components of
High Mass X-ray Binary (HMXB) systems with neutron stars undergoing
wind-fed accretion is discussed. The X-ray luminosity of these
systems allows us to evaluate the mass capture rate by the neutron
star from the stellar wind of its massive companion and set limits
on the relative velocity between the neutron star and the wind.
We find that the upper limit
to the wind velocity in the orbital plane during
the high state of the X-ray source is in the range 120-1000 km/s,
which is by a factor of 2--4 lower than both the terminal wind velocity
and the speed of the wind flowing out from the polar regions of massive stars
for all the objects under investigation. This finding is valid not
only for the systems with Be stars, but also for those system in which
the optical components do not exhibit the Be phenomenon. We also show
that the lower limit to the radial wind velocity in these systems
can unlikely be smaller than a few per cent of the orbital velocity
of the neutron star. This provide us with a new constraint on the
mass-transfer process in the outflowing disks of Be-type stars.}


\begin{document}


\citet{Delgado-Marti-etal-2000} highlighted that High Mass X-ray
Binaries (HMXBs) harboring neutron stars can, under certain conditions, serve as natural laboratories for studying stellar winds generated by early spectral-type stars. The neutron star in these systems can be considered as a probe. As it orbits through the stellar wind of its massive companion, the neutron star  captures gas via its gravitational field and accretes it onto its surface. The properties of the X-ray source produced by this process allow us to infer the amount of gas being captured by the neutron star in unit time. Combining this information with the observed parameters of the binary system and the characteristics of its massive component enable us to evaluate the physical conditions in the stellar wind emanating from the massive star in the orbital plane of the system.

Using their proposed methodology, \citet{Delgado-Marti-etal-2000}
analyzed the Be/X-ray binary system X\,Persei. They have found that the velocity of the wind which the Be star ejects within the orbital plane of the system is unlikely to exceed $150\,{\rm km\,s^{-1}}$. If the
velocity were higher, the amount of material captured by the neutron
star from the wind of its companion would be insufficient to explain
the observed X-ray luminosity of the pulsar. This result was not quite unexpected, since Be-type stars are known to be surrounded by the outflowing gaseous disk in which the radial velocity of material  is a factor of 3-5 smaller than the wind velocity in the polar regions \citep[see, for example,][and references therein]{Bradt-and-McClintock-1983}   and is almost an order of magnitude lower than the typical terminal wind velocity of hot
massive stars. This situation is just realized in X\,Persei, where the wind velocity of the Be star measured through observations of the UV lines,
 (about $800\,{\rm km\,s^{-1}}$) exceeds the velocity of the wind within the orbital plane by at least a factor of a few \citep{Delgado-Marti-etal-2000}.

An attempt to apply the method proposed in \citep{Delgado-Marti-etal-2000} to the HMXB OAO\,1657-415, in which the massive star does not show a Be-type phenomenon, has been made in our previous paper \citep{Ikhsanov-etal-2024}.  The object of our analysis was one of the best-known eclipsing HMXB systems containing a bright quasi-steady accretion-powered X-ray pulsar with orbital parameters and X-ray flux determined with high
confidence. The only uncertainty in its X-ray luminosity arises from
conflicting distance estimates to the source. It is
well-established, however, that the massive component does not fill
its Roche lobe, and mass transfer between the components occurs via
the wind-fed accretion scenario. Our derived upper limit for the wind
velocity of the massive star in OAO\,1657-415 ejected within the orbital plane is in the interval of $200-500\,{\rm km\,s^{-1}}$. It is higher than the wind velocity obtained by \citet{Delgado-Marti-etal-2000} for X\,Persei but remains well below both a typical terminal wind velocity of
early-type stars (about $2000\,{\rm km\,s^{-1}}$ for O-type stars \citep{Rodes-Roca-etal-2015, Kretschmar-etal-2019}  and $500-1500\,{\rm km\,s^{-1}}$ for B-type super-giants \citep{Kaper-etal-1995,Manousakis-and-Walter-2015})
and the wind velocity evaluated from the properties of UV lines observed in these objects \citep{Sadakane-etal-1985}. This raises a question about the structure of stellar wind of massive early-type stars which do not show the Be-phenomenon. In particular, is the velocity of wind ejected by these stars within the orbital plane smaller than the wind velocity in the polar regions?


In order to answer this question we consider in this paper a sample of HMXBs in which a neutron star undergoing wind-fed accretion is orbiting its massive companion which does not show Be-phenomenon. Measuring the
intensity of X-rays generated in this process allows us to estimate
the rate of gas accretion onto the surface of the neutron star and,
accordingly, the minimum rate at which it captures gas from the wind
of its companion. Incorporating this finding with observational data
on the massive component of the system, we arrive at an estimate of
the neutron star's velocity relative to the stellar wind of the
massive star, which is expelled in the orbital plane of the system
(see Section~\ref{base}). Using condition of the steady accretion process we also set a lower limit to the radial velocity of the outflowing matter (see Section~\ref{radial}). Following these methods we expand the list of
objects under study and present our estimates of the relative
velocity of the neutron star in some of the most thoroughly studied
HMXBs (see Section~\ref{rez}). The assumptions we made in our
analysis and some of the implications of the results presented are
briefly discussed in Section~\ref{discussion}.

\section{Upper limit to the relative velocity}\label{base}

The luminosity of a source arising due to the accretion of gas onto
the surface of a neutron star is estimated by the expression
\citep[see, for example,][and references therein]{Lipunov-1987}.
 \begin{equation}\label{lx}
 L_{\rm a} = \dot{M}_{\rm a} \frac{GM_{\rm ns}}{R_{\rm ns}}.
 \end{equation}
Here $\dot{M}_{\rm a}$ is the mass accretion rate, that is, the amount of
matter falling per unit time onto the surface of a neutron star,
whose mass and radius are, respectively, $M_{\rm ns}$ and $R_{\rm
ns}$. In the scenario of accretion onto a neutron star with a strong
magnetic field (which corresponds to the case of accretion we are
considering for X-ray pulsars), the energy of the accretion source
is predominantly emitted in the ``classical'' part of the X-ray
range (with photon energies $\sim 1-10$\,keV). This creates the most
favorable opportunity to evaluate the accretion rate onto the
surface of neutron stars,
 \begin{equation}\label{dmf-a}
 \dot{M}_{\rm a} = \frac{L_{\rm x} R_{\rm ns}}{GM_{\rm ns}},
 \end{equation}
through measurements of X-ray luminosity of the object, $L_{\rm x}$. A precise evaluation of $\dot{M}_{\rm a}$ requires only information about the distance to the source under investigation, while the mass and
radius of neutron stars are constrained within a relatively narrow
range of possible values ($R_{\rm ns}\sim 8 - 12$\,km and 
$M_{\rm ns}\sim 1 - 2\,M_{\odot}$) by their equation of state \citep{Potekhin-2010}. Because of the latter property, neutron stars prove to be a really good probe.

The rate at which a neutron star captures gas as it moves through
the stellar wind of its companion is estimated by the expression
  \begin{equation}\label{dmf-c}
\dot{M}_{\rm c} = \pi r_{\!_{\rm G}}^2 \rho_{\rm w} v_{\rm rel}
= \frac{4 \pi (GM_{\rm ns})^2 \rho_{\rm w}}{v_{\rm rel}^3},
 \end{equation}
where
  \begin{equation}
  r_{\!_{\rm G}} = \frac{2 GM_{\rm ns}}{v_{\rm rel}^2}
 \end{equation}
is a so-called Bondi radius \citep{Bondi-1952} which represents the maximum possible distance from the center of the neutron star to the point at which it is able to capture material as it moves through the stellar wind at the relative velocity
 \begin{equation}\label{v_rel_gen}
 v_{\rm rel} = \left(v_{\rm orb}^2 + v_{\rm w}^2
+ c_{\rm s(w)}^2\right)^{1/2}.
 \end{equation}
Here $v_{\rm orb}$ is the orbital velocity of the neutron star,
$v_{\rm w}$ is the stellar wind velocity in the reference frame of
the massive star, and $c_{\rm s(w)}$ is the sound speed in the
stellar wind. Finally, $\rho_{\rm w}$ is the density of the stellar
wind at the Bondi radius, which in the approximation of a spherically symmetric wind outflow can be expressed as
 \begin{equation}\label{rho-w}
 \rho_{\rm w} = \frac{\dot{M}_{\rm opt}}{4 \pi a^2 v_{\rm w}}.
 \end{equation}
Here, $\dot{M}_{\rm opt}$ is the mass-loss rate of the optical
component of the system (i.e. the massive star) in the form of the stellar wind, and $a$ is the orbital separation (distance between the system's components).

Combining expressions~\eqref{dmf-c} and \eqref{rho-w} under the
approximation $v_{\rm rel} \approx v_{\rm w}$ and solving the
inequality $\dot{M}_{\rm a} \leq \dot{M}_{\rm c}$ for $v_{\rm rel}$, we
find $v_{\rm rel} \leq v_{\rm max}$, where
  \begin{equation}\label{v-max}
 v_{\rm max} \simeq 650\,{\rm km\,s^{-1}} \times
L_{36}^{-1/4} R_6^{-1/4} m^{3/4} \left(\frac{a}{0.1\,{\rm
AU}}\right)^{-1/2}
\left(\frac{\dot{M}_{\rm opt}}
{10^{-7}\,{\rm M_{\odot}\,year^{-1}}}\right)^{ 1/4}
  \end{equation}
is the maximum possible value of the stellar wind velocity
in the orbital plane of a neutron star. Here $L_{36} = L_{\rm
x}/10^{36}\,{\rm erg\,s^{-1}}$, $R_6 = R_{\rm ns}/10^6$\,cm and $m =
M_{\rm ns}/1.4\,M_{\odot}$.

 \section{Lower limit to the wind radial velocity}\label{radial}

Studies of HMXBs also open the unique possibility of evaluating the minimum possible value of radial velocity of the stellar wind ejected by the massive star within the orbital plane of the system. The distance at which a neutron star in a HMXB system is able to capture gas as it moves through the wind of its massive companion in general case is limited to
  \begin{equation} \label{r-cap}
 r_{\rm cap} \leq \left\{
\begin{array}{lc}
r_{\!_{\rm G}}, \hspace{1mm} & {\rm for} \hspace{1mm}
v_{\rm rel} > v_0, \\
& \\
R_{\rm L_{\rm ns}}, \hspace{0mm}& {\rm for}
\hspace{1mm}
v_{\rm rel} \leq v_0, \\
\end{array}
\right.
 \end{equation}
where
  \begin{equation}\label{r-l-ns}
R_{\rm L_{\rm ns}} \simeq 3 \times 10^{11}\,{\rm cm} \, \left(\frac{a}{0.1\,{\rm AU}}\right)
 \end{equation}
is the radius of the Roche lobe of the neutron star and the velocity
  \begin{equation}
 v_0 \simeq 370\,{\rm km\,s^{-1}} \, m^{1/2} \, \left(\frac{a}{0.1\,{\rm AU}}\right)^{-1/2}
 \end{equation}
is defined by equation $r_{\!_{\rm G}}(v_0) = R_{\rm L_{\rm ns}}$ (for details see, e.g.
\citep{Ikhsanov-etal-2001} and references therein).
As a neutron star orbits its companion with the relative velocity $v_{\rm rel} \leq v_0$ it captures all material located inside the impact parameter, which in the case considered is equal to $R_{\rm L_{\rm ns}}$. A stationary accretion process can be realized in this case only if the distance in the radial direction which the stellar wind is able to pass on a timescale of the orbital period is comparable to or exceeds the scale of the impact parameter. Otherwise, the amount of gas located at the orbit of the neutron star would not be sufficient to produce observed luminosity of the X-ray pulsar. This condition implies that the stationary accretion can be realized if the radial velocity of the stellar wind is limited as $v_{\rm w(r)} \geq v_{\rm f}$, where 
 \begin{equation}
v_{\rm f} = \frac{R_{\rm L_{\rm ns}}}{P_{\rm orb}}
\simeq 1.2\,{\rm km\,s^{-1}} \times 
\left(\frac{ R_{\rm L_{\rm ns}}}{10^{11}\,{\rm cm}}\right) 
\left(\frac{P_{\rm orb}}{10\,{\rm d}}\right)^{-1}
\end{equation}
turns out to be in the interval
2--12\,km/s for all the systems presented in Table\,\ref{t1}.
This finding suggests that the radial velocity of stellar wind ejected by massive stars in HMXBs within the orbital plane is unlikely to be smaller than a few percent of the orbital velocity of the neutron star itself. This rather simple estimation should be, however, taken into account as a necessary condition in the modeling of outflowing disks surrounding the massive early-type stars.

Finally, the above method allows us to evaluate also the density of the wind by combining Eq.\,\eqref{dmf-c} with Eq.~\eqref{dmf-a} and solving it for $\rho_{\rm w}$.  The minimum possible value of the neutron star relative velocity can be estimated by its orbital velocity which in our case can be approximated by the keplerian velocity: $v_{\rm rel} \geq v_{\rm k}^{\rm (ns)}(a)$, where
  \begin{equation}\label{v_k}
v_{\rm k}^{\rm (ns)}(a) = \left(\frac{GM_2}{a}\right)^{1/2}\simeq
365\, \text{km\,s$^{-1}$} \times\left(\frac{M_2}{15\,M_{\odot}}\right)^{1/2}
\left(\frac{a}{0.1\,{\rm AU}}\right)^{-1/2},
 \end{equation}
and $M_2$ is the mass of the optical component of the system.
  Thus we get $\rho_{\rm w} \geq \rho_0$, where

 \begin{equation}\label{rho_0}
 \rho_0 = \frac{L_{\rm x} R_{\rm ns}}{4 \pi \left(GM_{\rm ns}\right)^3} \left(\frac{GM_2}{a}\right)^{3/2} \simeq 6 \times 10^{-16}\,\text{g\,cm$^{-3}$}\, L_{36} R_6 m^{-3} \left(\frac{M_2}{15\,M_{\odot}}\right)^{3/2}
\left(\frac{a}{0.1\,{\rm AU}}\right)^{-3/2}.
 \end{equation}

\section{Wind Velocity Estimation}
  \label{rez}

The results of the relative velocity evaluation for the neutron stars in the HMXBs,
using the methods described above and employing
the expression~\eqref{v-max} for the most thoroughly studied
persistent X-ray pulsars in the HMXBs are presented in the last column of 
Table\,\ref{t1}. It also provides the values of the key parameters of these
objects used in our calculations \citep[see][and references therein]{Falanga-etal-2015, Sidoli-Paizis-2018, Kim-etal-2023}, namely, the orbital period and size of the system, the spectroscopic mass, spectral class and mass-loss rate of the optical (massive) component and
the maximum luminosity of the X-ray source. 

\begin{table}[H]
\caption{Estimation of the maximum possible values of stellar wind velocity in the orbital plane of the binary system in the HMXBs along with their key parameters \citep[see][and references therein]{Falanga-etal-2015, Sidoli-Paizis-2018, Kim-etal-2023} used in our calculations (see text above for details).} \label{t1}
 \begin{adjustwidth}{-1cm}{-1cm}
    \begin{tabular}{l|c|c|c|c|c|c|c}
\toprule
    Object & $P_{\rm orb}$ & a & Massive star & $M_{\rm opt}$ &$\dot{M}_{\rm opt}$ &$L_{\rm x}^{\rm max}$&
     $\upsilon_{\rm max}$    \\
     & day & AU & &  $M_\odot$ &$10^{-7}\,M_{\odot}/{\rm yr}$ &$ 10^{36}$\,erg/s   & km/s \\
    \midrule
      Cen~X-3 &  2.03 & 0.08 & O6.5 II-III & 20 &50 &14 &   1010  \\
       4U~1538-52 & 3.73 & 0.11& B0.2 Ia & 20 & 12 & 4.3 &  810  \\
        Vela~X-1 & 8.96 & 0.23 & B0.5 Ia & 26 & 40 &0.10  &  610  \\
        OAO~1657-415 & 10.45 & 0.21 &  Ofpe/WNL & 14  & 1&20 &   210  \\
        2S~0114+650 & 11.6 & 0.26 & B1 Ia &16 & 30 &12 &    510  \\
        GX~301-2 & 41.5 & 0.83 &B1.5 Ia &43 & 75 &30 &    290   \\
        X~Per & 250 & 2.0 & B0 Ve &15& 0.05  &0.06  &  140  \\
        RX~J0146.9+6121 & 330 & 2.1& B1 Ve &10&0.05 &0.11 &    120 \\

\bottomrule
\end{tabular}
	\end{adjustwidth}
\end{table}

\section{Discussion}\label{discussion}

The estimates we obtained for the upper limit
to the relative velocity between the neutron star and stellar wind
of its companion indicate that the wind speed in the orbital plane
of the system is lower than the typical terminal wind
velocity of O-type stars by a factor of 2-4 and is comparable with the expected wind velocity for B-type supergiants. Estimated  velocity of the wind ejected in the orbital plane of two Be/X-ray pulsars 
is smaller than the wind velocity measured through observations of the UV lines in the spectra of massive stars by a factor of a few.

This result supports the hypothesis about a multi-component
structure of the stellar wind of massive stars. It should be noted
that the moderate wind velocity in the orbital plane of the system
turns out to be inherent not only to the systems with Be stars, for
which this result is rather expected, but also to the systems with
optical components, which do not show the Be phenomenon and/or are
the stars of a significantly earlier spectral class.

The obtained values of relative velocity in the systems with short orbital periods (compact systems) are somewhat higher than in the long-period binaries (wide
pairs). Moreover, there is a noticeable tendency
towards a decrease in the upper limit to the relative velocity as the system size increases. This may indicate that the velocity of the wind flowing out in the orbital plane of the system changes with the distance from the massive component insignificantly.

Finally, the obtained lower limit to the radial velocity of the stellar
wind, $v_{\rm f} \simeq 2-12\,{\rm km\,s^{-1}}$, is comparable to the sound speed in the gas heated to a few thousand
Kelvin and is unlikely to be smaller
than a few per cent of the orbital velocity of the neutron star
itself. This finding challenges scenarios in which the mass
outflow from a Be star is treated in terms of a hydrodynamical cool
moderately viscous disk. Incorporation of the magnetic field of
the stellar wind into the model may help to increase the efficiency
of the mass and angular momentum transfer out from the star and to
explain a relatively high value of the lower limit to the radial velocity
of the wind in the orbital plane, $v_{\rm f}$.

\authorcontributions{N.I. --- conceptualization and methodology, writing---review and editing; N.B. --- validation, writing---original draft preparation and editing, V.K. --- data curation and investigation.  All authors have read and agreed to the published version of the manuscript.}

\funding{This research was carried out with  the financial support of  the Ministry of Science and Higher Education of the Russian Federation, grant No. 075-15-2022-262 (13.MNPMU.21.0003), within the framework of the program for the study of massive stars with the 6-m telescope of SAO RAS (BTA).
}

\acknowledgments{We would like to thank all the referees for useful comments. NB and NI express their gratitude  to the Fesenkov Astrophysical Institute for fruitful
cooperation and warm hospitality and to the Ministry of Science and Higher Education of the Russian Federation for the support of this study within the national project "Science and universities''.}

\conflictsofinterest{The authors declare no conflicts of interest.}



\isPreprints{}{
\begin{adjustwidth}{-\extralength}{0cm}
} 
\printendnotes[custom] 

\reftitle{References}



\begin{thebibliography}{999}
\bibitem[Delgado-Marti et~al.(2000)]
{Delgado-Marti-etal-2000}
Delgado-Mart\'i, H.; Levine, A.M.; Pfahl, E.; and Rappaport, S.A.
The orbit of X~Persei and its neutron star companion.
\apj\ {\bf 2000}, {\em 546}, 455-469.

\bibitem[Bradt and McClintock (1983)]{Bradt-and-McClintock-1983}
Bradt, H.V.D.; McClintock, J.E.Bradt, H.V.D.; McClintock, J.E. The Optical Counterparts of Compact Discrete Galactic X-Ray Sources.
Annual Rev. Astron. Astrophys.{\bf 1983} {\em 21}, 13-66


 \bibitem[Ikhsanov et~al.(2024)]
{Ikhsanov-etal-2024} Ikhsanov, N.R.; Kim, V.Y.; Beskrovnaya, N.G.,
On the estimate of the stellar wind velocity of the massive component in OAO\,1657-415.
\textit{Publ. of Pulkovo Observatory} {\bf 2024}, {\em 233}, 34-38.


\bibitem[Rodes-Roca et~al. (2015)]{Rodes-Roca-etal-2015} Rodes-Roca, J. J.; Mihara, T.; Nakahira, S. et al., Orbital phase-resolved spectroscopy of 4U 1538-52 with MAXI.
\aap\ {\bf 2015} {\em 580}, A140,5p.

\bibitem[Kretschmar et~al. (2019)]{Kretschmar-etal-2019} Kretschmar, P.; Martinez-Nunez, S.; Furst, F. et al., Vela X-1 as a laboratory for accretion in High-Mass X-ray Binaries.
 Mem. Soc. Astron. Italiana {\bf 2019} {\em 90}, 221-224

\bibitem[Kaper et~al. (1995)]{Kaper-etal-1995} Kaper, L.; Lamers, H. J. G. L. M.; Ruymaekers, E. et al.  Wray 977 (GX 301-2): a hypergiant with pulsar companion.
\aap {\bf 1995} {\em 300}, 446-452

\bibitem[Manousakis and Walter (2015)]{Manousakis-and-Walter-2015} Manousakis, A., Walter, R.   The stellar wind velocity field of HD 77581.
\aap {\bf 2015} {\em 584}, A25, 5p.

\bibitem[Sadakane et~al.(1985)]
{Sadakane-etal-1985}
Sadakane, K.; Hirata, R.; Jugaku, J.; et al.
Ultraviolet Spectroscopic Observations of HD 77581 (VELA X-l = 4U 0900-40).
\apj\ {\bf 1985}, {\em 288}, 284-291.

\bibitem[Lipunov(987)]{Lipunov-1987}
Lipunov, V.M. {\textit Astrophysics of Neutron Stars},
Moscow, Nauka, 1987.

\bibitem[Potekhin(2010)]{Potekhin-2010}
Poteknin, A.Yu. Physics of neutron stars, Uspekhi {\bf 2010} {\em 180}, 1279-1304.

\bibitem[Bondi(1952)]{Bondi-1952}
Bondi, H.  On spherically symmetrical accretion \mnras\ {\bf 1952} {\em 112}, 195-204.

\bibitem[Ikhsanov et~al. (2001)]{Ikhsanov-etal-2001}
Ikhsanov, N.R.; Larionov, V.M.; Beskrovnaya, N.G.
On the accretion flow geometry in A0535+26. \aap\ {\bf 2001} {\em 372}, 227-232.

\bibitem[Falanga et~al.(2015)]{Falanga-etal-2015}
Falanga, M.; Bozzo, E.; Lutovinov, A. et~al.
Ephemeris, orbital decay, and masses of ten eclipsing high-mass
 X-ray binaries. \aap\ {\bf 2015} {\em 577}, A130, 1-16.

\bibitem[Kim et~al.(2023)]{Kim-etal-2023}
Kim, V.;  Izmailova, I.; Aimuratov, Y. (2023), Catalog of the Galactic Population of X-Ray Pulsars in High-mass X-Ray Binary Systems.
\apjs\ {\bf 2023} {\em 268}, (1), 21.

\bibitem[Schulz et~al.(2003)]{Schultz-etal-2003}
Schulz, N.S.; Canizares, C.; Huenemoerder, D.; Tibbets, X-ray modeling of very young early-type stars in the orion trapezium: signatures of magnetically confined plasmas and evolutionary implications. \apj\ {\bf 2003} {\em 595}, 365-383.

\bibitem[Sidoli and Paizis(2018)]{Sidoli-Paizis-2018}
Sidoli, L.; Paizis, A.  An INTEGRAL overview of High-Mass X-ray Binaries: classes or transitions? \mnras\ {\bf 2018} {\em 481}, 2779-2803.


\end{thebibliography}


\isAPAandChicago{}{

}{}

\PublishersNote{}
\end{adjustwidth}
\end{document}